\documentclass[twocolumn,amsmath,aps]{revtex4}

\usepackage{amssymb,mathrsfs,bm,color,times}
\usepackage{graphicx,color}
\usepackage{CJK}
\usepackage{bm}
\usepackage[bookmarks=false]{hyperref}
\usepackage{amsmath}
\usepackage{subfigure}
\hypersetup{colorlinks=true,citecolor=blue,linkcolor=blue,urlcolor=blue,pdfstartview=FitH,bookmarksopen=true}

\newcommand{\be}{\begin{equation}}
\newcommand{\ee}{\end{equation}}
\newcommand{\bea}{\begin{eqnarray}}
\newcommand{\eea}{\end{eqnarray}}
\newcommand{\bsube}{\begin{subequations}}
\newcommand{\esube}{\end{subequations}}

\newcommand{\Eq}[1]{Eq.\,(\ref{#1})}
\newcommand{\Eqs}[1]{Eqs.\,(\ref{#1})}

\newcommand{\dg}{\dagger}
\newcommand{\la}{\langle}
\newcommand{\ra}{\rangle}

\newcommand{\nl}{\nonumber \\}

\usepackage{amsfonts}
\usepackage{amsmath}
\usepackage{amssymb}
\usepackage{graphicx}
\usepackage{type1cm}
\usepackage{times}
\usepackage{bm}
\usepackage{color}

\newcommand{\beq}{\begin{equation}}
\newcommand{\eeq}{\end{equation}}
\newcommand{\beqn}{\begin{eqnarray}}
\newcommand{\eeqn}{\end{eqnarray}}
\newcommand{\bsub}{\begin{subequations}}
\newcommand{\esub}{\end{subequations}}

\begin{document}
\begin{CJK}{GBK}{song}

\title{Postselected amplification applied to Mach-Zehnder-interferometer
for phase shift measurement of optical coherent states}

\author{Jialin Li}
\affiliation{Center for Joint Quantum Studies and Department of Physics,
School of Science, \\ Tianjin University, Tianjin 300072, China}
\author{Yazhi Niu}
\affiliation{Center for Joint Quantum Studies and Department of Physics,
School of Science, \\ Tianjin University, Tianjin 300072, China}

\author{Lupei Qin }
\email{qinlupei@tju.edu.cn}
\affiliation{Center for Joint Quantum Studies and Department of Physics,
School of Science, \\ Tianjin University, Tianjin 300072, China}

\author{Xin-Qi Li}
\email{xinqi.li@imu.edu.cn}
\affiliation{Center for Quantum Physics and Technologies,
School of Physical Science and Technology,
Inner Mongolia University, Hohhot 010021, China}

\date{\today}

\begin{abstract}
We propose a postselected amplification (PSA) scheme for phase shift measurement
of optical coherent states when passing through
the Mach-Zehnder-interferometer (MZI).
Different from the usual weak-value-amplification (WVA) formulation,
the which-path states of the MZI ($|1\ra$ and $|2\ra$)
cannot be described as sub-system states
entangled with the optical coherent states
($|\alpha_1\ra$ and $|\alpha_2\ra$) separated by the beam-splitter.
However, we obtain the same result of the usual WVA
in the Aharonov-Albert-Vaidman (AAV) limit,
but beyond this limit, the result is different,
e.g., the photon-number scaling can be different.
We explicitly carry out the amplified phase shift,
which is extracted out from
the field-quadrature measurement in the dark port of the MZI.
We also evaluate the performance quality of the proposed scheme,
and analyze the technical advantages
by considering possible errors in the quadrature measurement.
\end{abstract}


\maketitle

{\flushleft\it Introduction}.---
Various laser-interferometers perform important roles in quantum precision measurements.
Physically, the laser beam is an optical coherent state ($|\alpha\ra$),
which is not a quantum ensemble of single photons.
This implies that, when the laser beam enters the interferometer
through, e.g., a beam splitter (BS),
the resultant state {\it cannot} be described as a quantum superposition
of two coherent states, such as $c_1|1\ra|\alpha_1\ra+c_2|2\ra|\alpha_2\ra$,
where $|\alpha_1\ra$ and $|\alpha_2\ra$ are determined by the beam splitter,
and $|1\ra$ and $|2\ra$ are the path-information states of the interferometer.
Actually, the optical field after the action of the beam splitter
should be described as a product state of $|\alpha_1\ra$ and $|\alpha_2\ra$.

However, in some type of measurements,
e.g., based on deflection of transverse spatial wavefunction $\Phi(x,y)$
(suppose the light propagating along the $z$-direction)
and polarization (or path-information states) \cite{Kwi08,How09a,How09b,How10a,How10b},
the result can be interpreted by a description of ensemble of single photons.
The reason is that in these measurements, all the photons in $|\alpha\ra$
have the same $\Phi(x,y)$ and polarization (or path-information states).
Nevertheless, for projective photon-number or field-quadrature measurement,
the description of {\it quantum ensemble of single photons} does not work.

\begin{figure}
  \centering
  \includegraphics[scale=0.5]{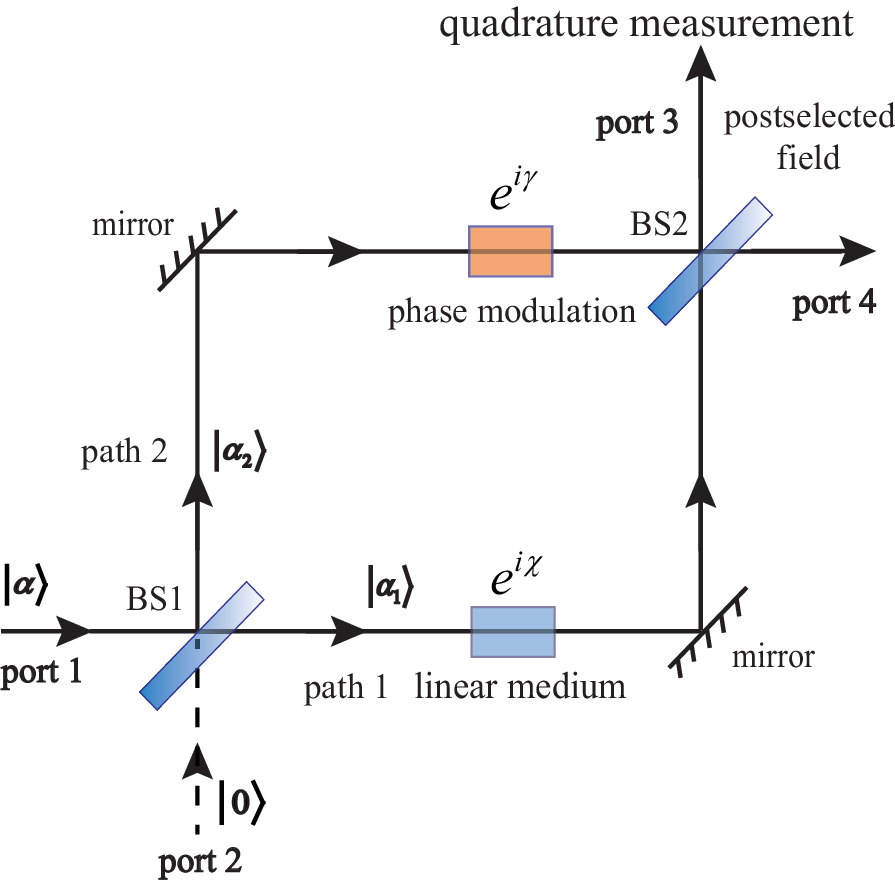}
  \caption{
Phase shift measurement of an optical coherent state $|\alpha\ra$,
when passing through the Mach-Zehnder-interferometer (MZI).
The phase shift $\chi$ is caused by interaction of $|\alpha\ra$
with a piece of linear optics medium,
while its extraction is via a homodyne field-quadrature measurement
to the postselected photons, exiting from the port 3.
The postselection can be realized by modulating
either the interference phase, as indicated in the upper arm,
or the transmission-and-reflection coefficients of the second beam splitter (BS2).
For a reason that is explained in the main text,
in our proposal, the second choice is suggested.   
Notice that
the optical coherent state $|\alpha\ra$ is split by the first beam splitter (BS1)
into a product state of $|\alpha_1\ra$ and $|\alpha_2\ra$,
but not their quantum superposition.
This implies that $|\alpha_1\ra$ and $|\alpha_2\ra$
cannot quantum mechanically entangle
with the path-information states $|1\ra$ and $|2\ra$.
As a consequence, the postselection does not fall into
the standard formulation of WVA by Aharonov, Albert, and Vaidman (AAV).}
\end{figure}

In this work, we consider the {\it phase shift} measurement of a laser beam
passing through a Mach-Zehnder-interferometer (MZI).

This setup and the phase shift measurement have been studied
in connection with probing relativistic gravity effects,
such as estimating the characteristic parameters
of black-holes and wormholes \cite{Ra17,Sab17,Sab18}.  
In this work, beyond Refs.\ \cite{Ra17,Sab17,Sab18},  
we propose a strategy of postselection
to amplify the phase shift \cite{AAV1,AAV2,Llo20,Shu22,Yang23},
which is extracted from
the field-quadrature measurement in the dark port of the MZI.
For a laser beam entering the MZI, one may feel the theoretical treatment is similar to
that of a two-state (qubit) system coupled to an optical coherent state $|\alpha\ra$,
as studied in Refs.\ \cite{ZLJ15,Aha15,XQLi22,XQLi24},
by an analogy that the which-path states $|1\ra$ and $|2\ra$ in the MZI
just corresponds to the qubit states $|g\ra$ and $|e\ra$.
In that case, for an initial qubit state $c_1|g\ra +c_2|e\ra$,
the total entangled state is $c_1|g\ra|\alpha_g\ra+c_2|e\ra|\alpha_e\ra$,
after the coupling interaction.
However, for the setup of MZI,
owing to the difference explained above (in the first paragraph),
the theoretical treatment of the postselected amplification (PSA) scheme
does not fall into the standard formulation
of the weak-value-amplification (WVA) \cite{AAV1,AAV2}.
Despite the difference we have pointed out, we will show that the same result
of the usual WVA in the Aharonov-Albert-Vaidman (AAV) limit can be obtained.
Nevertheless, beyond the AAV limit, the result is different,
e.g., the photon-number scaling is essentially different \cite{XQLi22,XQLi24}.
We will explicitly carry out the amplified phase shift
in the postselected photons, which can be extracted from
the field-quadrature measurement in the dark port of the MZI.
We will also analyze the performance quality of the PSA
and illustrate the technical advantages.
Rather than the shot noise,
which is the main limitation in quantum-limited measurement,
we will consider possible errors in the quadrature measurement.

\vspace{0.1cm}
{\flushleft\it Theoretical Formulation of the PSA}.---
Let us consider a laser beam entering the MZI through a beam splitter, as shown in Fig.\ 1.
To provide a quantum mechanical description (in the Schr\"odinger picture)
for the propagation of the coherent state $|\alpha\ra$ through the MZI,
we first outline the simple theory for the beam splitter.
One can regard the laser beam as a classical wave.
Then, the outgoing waves ($c$ and $d$)
are related with the incident waves ($a$ and $b$)
through a scattering matrix, as follows
\begin{equation}
\begin{aligned}
\left( {\begin{array}{*{20}{c}}
c\\
d
\end{array}} \right) = S \left( {\begin{array}{*{20}{c}}
a\\
b
\end{array}} \right)  \,.
\label{matrix}
\end{aligned}
\end{equation}
The scattering matrix can be parameterized as
\bea
S = \left( {\begin{array}{*{20}{c}}
t&r\\
r&t
\end{array}} \right) = \left( {\begin{array}{*{20}{c}}
{\cos \theta }&{ - i\sin \theta }\\
{ -i\sin \theta }&{\cos \theta }
\end{array}} \right) \,,
\eea
with $t$ and $r$ the transmission and reflection coefficients,
satisfying ${\left| t \right|^2} + {\left| r \right|^2} = 1$.
Corresponding to this,
the quantum state transformation by the BS1 of the MZI is given by
\bea
U_{BS1}\left| {\alpha ;0} \right\rangle
= \left| {\alpha \cos \theta_1 ; - i\alpha \sin \theta_1 } \right\rangle  \,,
\eea
where the unitary transformation operator reads as
$U_{BS1} = e^{ - i\theta_1 ({{\hat a}^\dag }{\hat b} + \hat a{{\hat b}^\dag })}$,
with the photon creation (annihilation) operator $\hat{a}^{\dg}$ ($\hat{a}$)
acting on the optical filed entering from the horizontal direction,
and the creation (annihilation) operator $\hat{b}^{\dg}$ ($\hat b$)
on the filed entering from the vertical direction.
This convention implies that,
in the state $|\alpha \cos \theta_1 ; - i\alpha \sin \theta_1\ra$,
the field state
propagating in the lower horizontal arm in Fig.\ 1
is $|\alpha_1\ra = |\alpha \cos \theta_1\ra$,
and the state propagating in the left vertical arm
is $|\alpha_2\ra = |- i\alpha \sin \theta_1\ra$.

After the action of BS1, the light propagating in the lower arm (path 1)
passes through a linear optical medium
and gains an extra phase shift,
described by $U_1 = e^{ i\chi  {\hat a}^\dag \hat a }$
acting on the coherent state $|\alpha_1\ra$.
$\chi$ is the interaction strength, which is the parameter to be estimated.
The light propagating in the other path (path 2)
can be modulated by a phase shifter,
described by $U_2 = e^{ i\gamma  {\hat b}^\dag \hat b }$
acting on the state $|\alpha_2\ra$, with $\gamma$ the phase modulation parameter.
In combination, the joint propagation of the incident laser beam
(coherent state $|\alpha\ra$) through the MZI is given by
\begin{equation}
\begin{aligned}
\left| {{\psi _3};{\psi _4}} \right\rangle
&= {U_{BS2}}{U_2}{U_1}{U_{BS1}}\left| {\alpha; 0} \right\rangle  \\
&= {\left| {{\alpha _{{f}}}} \right\rangle _3}{\left| {{{\alpha _{\bar f}}}} \right\rangle _4}
\label{psi34}
\end{aligned}
\end{equation}
with
\begin{eqnarray}
\label{alf}
{\alpha _f} &=& \frac{\alpha }{{\sqrt 2 }}({e^{i\chi }}
\cos {\theta _2} - {e^{i\gamma }}\sin {\theta _2})\label{alphaf},  \nl
{{\alpha _{\bar f}}} &=&  - i\frac{\alpha }{{\sqrt 2 }}({e^{i\chi }}
\sin {\theta _2} + {e^{i\gamma }}\cos {\theta _2})\label{alphafbar}.
\end{eqnarray}
Here we used $|\psi_3\ra$ and $|\psi_4\ra$
(and also $|\alpha_f\ra_3$ and $|\alpha_{\bar f}\ra_4$)
to denote the field states exiting from the port 3 and port 4, respectively.

In obtaining this result, we assume $\theta_1=\frac{\pi}{4}$ for BS1,
and remain $\theta_2$ to be changeable for BS2
(for a reason to be explained later, in the paragraph below \Eq{chi-A}).

In this context, we may mention that the conventional measurement
can employ the difference of the light intensities in port 3 and port 4,
i.e., $\delta I=I_3-I_4$,
to extract the phase shift $\chi$ (after a procedure of pre-calibration)
from the interference pattern by modulating the phase shift parameter $\gamma$,
while the pattern
($\delta I$ {\it versus} $\gamma$)
is just like the double-slit interference pattern of intensity
as a function of position.

However, in our PSA scheme, we will consider to exact the phase shift $\chi$
from the field quadrature measurement
in one of the exit ports, e.g., in port 3,
as considered in Refs.\ \cite{Ra17,Sab17,Sab18}.
As will be clear later, for this purpose,
modulating the transmission-reflection parameter $\theta_2$ of BS2
can be a proper choice.

Now let us return to the results of \Eqs{psi34} and (\ref{alf})
and regard the outgoing field from port 3 as the postselected result.
Introducing $c_1=\cos\theta_2/\sqrt{2}$
and $c_2=-e^{i\gamma}\sin\theta_2/\sqrt{2}$,
we can reexpress $\alpha_f$ as
\bea\label{alf-2}
\alpha_f &=& (c_1 e^{i\chi} + c_2)\alpha   \nl
&\simeq&  [c_1 (1+i\chi)  +c_2] \alpha    \nl
&=& (c_1+c_2) \left( 1+i\chi \frac{c_1}{c_1+c_2}  \right) \alpha  \nl
&\simeq& (c_1+c_2) e^{i A_w \chi} \alpha   \,.
\eea
Here we introduced $A_w = c_1 /(c_1+c_2)$.
Indeed, this result can be identified as the AAV's weak value, by noting that
\bea
A_w = \frac{c_1}{c_1+c_2} = \frac{\la f|\hat{A}|i\ra}{\la f|i\ra}  \,,
\eea
if we recognize that $\hat{A}=|1\ra \la 1|$,
and the pre- and post-selected path-information states are
$|i\ra = (|1\ra -i |2\ra)/\sqrt{2}$,
and $|f\ra = \cos\theta_2|1\ra +i \sin\theta_2 e^{-i\gamma}|2\ra)/\sqrt{2}$.

Then, from the last line of \Eq{alf-2}, we find that
the phase shift $\chi$ is amplified as
\bea\label{chi-A}
\widetilde{\chi}_A=A_w \chi  \,.
\eea
This is the standard result of WVA in the AAV limit.
Performing a homodyne quadrature measurement for the optical field in this port,
one can extract the amplified phase shift $\widetilde{\chi}_A$,
which is more tolerant to technical imperfections
than $\chi$ (in particular for small $\chi$).

The amplification effect can be understood as follows.
Notice that $c_1+c_2=(\cos\theta_2 - e^{i\gamma}\sin\theta_2)/\sqrt{2}$.
If setting $\gamma=0$,
i.e., not introducing the element of phase shifter in the MZI,
then making $\theta_2=\pi/4$ would result in port 3
being a completely dark port (no light is going out from this port).
Further, one can modulate the transmission-reflection coefficient of BS2.
Making $\theta_2$ slightly deviate from $\pi/4$,
one can realize the amplification, since $c_1+c_2$ is small, thus $A_w$ is large.

One may consider a different choice, say, setting $\theta_2=\pi/4$
and modulating the phase shift $\gamma$.
In this case, $c_1+c_2 =(1-e^{i\gamma})/2$.
For a small $\gamma$, $c_1+c_2$ can be small,
but is imaginary, i.e., $c_1+c_2\simeq -i\gamma/2$.
Then, $e^{i\widetilde{\chi}_A}$ is not a phase shift
being added to the phase of $\alpha$.
This factor cannot be extracted from the field quadrature measurement.
Another point is that $|\alpha_f|^2=|c_1+c_2|^2 N$, with $N=|\alpha|^2$.
This implies that the intensity of the outgoing light is weak.
Indeed, this is the common feature of all WVA and PSA schemes.
However, as to be analyzed later, in some cases,
this is not a serious problem.

The approximated treatment of \Eq{alf-2} is valid in the so-called AAV limit.
Beyond this limit, one can reexpress the result of $\alpha_f$ as
 \begin{equation}
 \label{alf-3}
\begin{aligned}
{\alpha_f} = \left| \alpha_f \right|{e^{i(\widetilde{\chi}_B + \lambda )}}  \,,
\end{aligned}
\end{equation}
where
\begin{eqnarray}\label{chi-B}
\left|\alpha_f \right| &=& \sqrt {\frac{N}{2}}
\sqrt {1 - \sin (2{\theta _2})\cos (\chi-\gamma) } \,,  \nl
\widetilde{\chi}_B &=& {\rm{arctan}}\left(\frac{{\sin \chi \cos {\theta _2}
- \sin \gamma \sin {\theta _2}}}{{\cos \chi \cos {\theta _2}
- \cos \gamma \sin {\theta _2}}} \right)  .
\end{eqnarray}
Just like $\widetilde{\chi}_A$, $\widetilde{\chi}_B$ is the amplified phase shift.
Once $\widetilde{\chi}_B$ is measured out,
the true phase shift $\chi$ can be extracted, from \Eq{chi-B}.
In Fig.\ 2, we numerically illustrate the behaviors of $\widetilde{\chi}_A$
and $\widetilde{\chi}_B$ {\it versus} $\theta_2$,
for $\chi=10^{-4}$ (AAV limit) and $10^{-2}$ (beyond),
with both behaviors resembling each other,
commonly showing the postselection induced amplification effect.

The present setup of phase shift measurement based on MZI
looks quite similar to a two-state (qubit) system
coupled to an optical coherent state $|\alpha\ra$,
as studied in Refs.\ \cite{ZLJ15,Aha15,XQLi22,XQLi24},
by regarding the which-path states $|1\ra$ and $|2\ra$ in the MZI
as the qubit states $|g\ra$ and $|e\ra$.
However, in that case, for an initial qubit state $c_1|g\ra +c_2|e\ra$,
the total entangled state is $c_1|g\ra|\alpha_g\ra+c_2|e\ra|\alpha_e\ra$,
after the coupling interaction.
After postselection, the resultant meter state (the optical field state)
is a superposition of two coherent states,
i.e., $|\widetilde{\Phi}_f \ra \sim \tilde{c}_1|\alpha_g\ra+ \tilde{c}_2|\alpha_e\ra$,
with $\tilde{c}_1=\la f|1\ra c_1$ and $\tilde{c}_2=\la f|2\ra c_2$,
while $|f\ra$ is the postselection state of the qubit.
This is actually a cat-state.
In the AAV limit, one can prove \cite{XQLi22,XQLi24} that
$|\widetilde{\Phi}_f\ra \sim |\alpha e^{i\widetilde{\chi}_A}\ra$,
which is the standard WVA result,
with the phase shift amplified from $\chi$ to $\widetilde{\chi}_A=A_w\chi$.
However, beyond the AAV limit,
the highly quantum nature of superposition
of two coherent states in $|\widetilde{\Phi}_f \ra$
can result in enhancement of the $N$ scaling behavior,
e.g., from $1/\sqrt{N}$ to $1/N$ \cite{XQLi22,XQLi24},
where $N$ is the average photon number of the probing field state $|\alpha\ra$.
In contrast, for the case of MZI, even beyond the AAV limit,
the optical field is just a single coherent state $|\alpha_f\ra$,
but not a quantum superposition of two coherent states,
with $\alpha_f$ given by \Eq{alf-3},
which can only result in the $1/\sqrt{N}$ scaling.
Nevertheless, as shown in Fig.\ 2, the desired PSA effect
can be induced by postselection.

\begin{figure}
  \centering
  \includegraphics[scale=0.5]{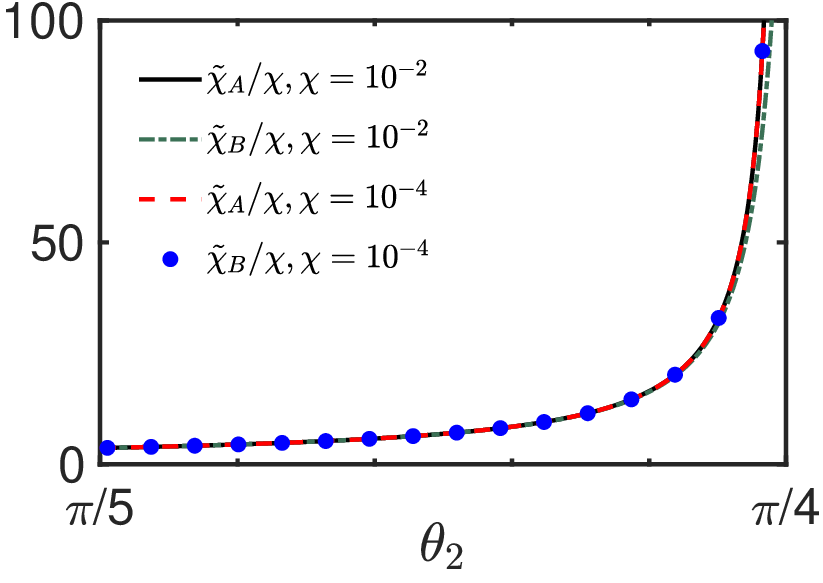}
  \caption{
Phase shift amplification in the postselected photons,
from the original $\chi$ to $\widetilde{\chi}$,
{\it versus} the postselection parameter $\theta_2$,
which characterizes the transmission-and-reflection of BS2 (see Fig.\ 1).
The results $\widetilde{\chi}_A$ and $\widetilde{\chi}_B$ [cf. eq 8 and eq 10]
are plotted for $\chi=10^{-4}$ and $10^{-2}$,
respectively, for comparison.
For the larger $\chi=10^{-2}$, difference is observed
between $\widetilde{\chi}_B$ and $\widetilde{\chi}_A$,
despite the qualitatively similar amplification behavior.      }
\end{figure}

\vspace{0.1cm}
{\flushleft\it Field Quadrature Measurement}.---
In order to extract the phase of an optical coherent state via measurement,
an efficient method is performing the homodyne measurement
for the field quadrature \cite{WM09}.
In this type of measurement, a reference light (the so-called LO light \cite{WM09}),
$\beta=|\beta |e^{i\xi}$, is introduced to mix with the signal light,
$\alpha=|\alpha| e^{i\lambda}$, through a beam splitter.
Then, from the current difference of two photo-detectors and
rescaling it with the strength of the reference light, $2|\beta|$,
one can obtain $\overline{X}_{\xi}=|\alpha|\cos(\lambda-\xi)$,
which is the quantum average of the quadrature operator
$\hat X_{\xi} = \frac{1}{2}({{\hat a}^\dag }{e^{i\xi }} + \hat a{e^{ - i\xi }})$
in the signal state $|\alpha\ra$, with $\xi$ the phase of the reference light.

Now consider the homodyne measurement of the exiting light in the dark port,
say, the coherent state $|\alpha_f\ra$.
In the AAV limit, the quadrature (quantum average) result is
\bea\label{XA}
\overline{X}_{\xi,\rm AAV}
&=& \sqrt {\frac{N}{2}}(\cos {\theta _2} - \sin {\theta _2})\sin\widetilde{\chi}_A  \,,
\eea
while beyond the AAV limit, the result is
\bea\label{XB}
\overline{X}_{\xi,\rm Byd}
&=& \sqrt {\frac{N}{2}} \sqrt {1 - \sin (2{\theta _2})\cos \chi } \sin\widetilde{\chi}_B \,.
\eea
In obtaining this result, we have noticed that the phase of $\alpha_f$
is $\lambda+\widetilde{\chi}_A$ in the AAV limit,
and is $\lambda+ \widetilde{\chi}_B$ when beyond the AAV limit.
We have also assumed
the phase of the reference light as $\xi=\pi/2 + \lambda$.
This choice makes the quadrature measurement result
most sensitive to the phase under estimation,
in the above, which is $\widetilde{\chi}_A$ or $\widetilde{\chi}_B$.

The above results, say, \Eqs{XA} and (\ref{XB}),
correspond to the following quantum ensemble average
\bea\label{XAB}
\overline{X}_{\xi}=\la\alpha_f|\hat{X}_{\xi}|\alpha_f\ra
= \sqrt{I_f}\sin\widetilde{\chi} \,.
\eea
Here and in some places in the following,
we use $\overline{X}_{\xi}$ to denote in a unified way
the result of quadrature measurement,
for both in the AAV limit and beyond it.
Accordingly, we get rid of the subscripts `A' and `B',
which indicate the AAV limit and beyond it.
In \Eq{XAB}, we have replaced
the theoretical result of the quadrature amplitude in \Eq{XA} or (\ref{XB})
with the one obtainable in experiment, i.e., $\sqrt{I_f}$,
while $I_f$ can be obtained from $I_f=(I_1+I_2)-I_{\rm LO}$.
In this simple equality, $I_1$ and $I_2$ are the light intensities
measured by the two photo-detectors in the homodyne measurement,
and $I_{\rm LO}$ is the intensity of the the LO light (i.e., $I_{\rm LO}=|\beta|^2$).

In quantum mechanics, with respect to the ensemble average of quantum measurements,
the random output result of each single measurement will deviate from it.
The deviation is statistically characterized
by the quantum fluctuation/uncertaity,
say, $\delta X_{\xi} = \sqrt{\la\hat{X}_{\xi}^2 \ra-(\bar{X}_{\xi})^2}$,
in the problem under present consideration,
with $\la \bullet \ra$ the quantum average in the state $|\alpha_f\ra$.
Then, one can define the signal-to-noise ratio (SNR)
of the quadrature measurement as $R^{S/N}=\overline{X}_{\xi}/\delta X_{\xi}$.
Moreover, using the error propagation formula,
for instance, for the case under the AAV limit,
$\delta\widetilde{\chi}
= \delta X_{\xi,\rm AAV} \, |\frac{\partial \overline{X}_{\xi,\rm AAV}}
{\partial \widetilde{\chi}}|^{-1} $,
one can define the {\it sensitivity}
of precision measurement (parameter estimation) through
$\widetilde{R}^{S/N}_{\rm AAV} = \widetilde{\chi}_A / \delta\widetilde{\chi}_A$.
After simple algebra, one can obtain
\bea
\widetilde{R}^{S/N}_{\rm AAV}
=\sqrt {2N} \left| {(\cos {\theta _2}
- \sin {\theta _2})\cos \widetilde \chi_A } \right|\, \widetilde \chi_A  \,.
\eea
Similarly, beyond the AAV limit, one can define
$\widetilde{R}^{S/N}_{\rm Byd} = \widetilde{\chi}_B/ \delta\widetilde{\chi}_B$,
and obtain
\bea
\widetilde{R}^{S/N}_{\rm Byd}
= \sqrt {2N[1 - \sin (2{\theta _2})\cos \chi ]} \,
\left| \cos \widetilde{\chi}_B   \right|\, \widetilde{\chi}_B   \,.
\eea

The above characterization in terms of SNR or sensitivity
corresponds to the fluctuation of {\it each single measurement result},
with respect to the quantum ensemble average
(theoretical average over infinite times of measurement).
However, in practice, one should use the average result
of many times (e.g.,$M$ times) measurement,
rather than a single time measurement result, to estimate the parameter.
For the problem under study,
the average result of quadrature measurement of $M$ times
can be expressed as
\bea
X_{\xi}^{(M)} = \frac{1}{M} \sum_{j=1}^M X_{\xi,j} \,,
\eea
where $X_{\xi,j}$ is the measurement result of the $j_{\rm th}$ time.
Conceptually speaking, $M$ times measurement considered here
corresponds to using $M$ laser pulses,
while each pulse is described by the coherent state $|\alpha\ra$.
In real experiment, this corresponds to using a continuous laser beam
and making average of the measurement result
over certain period of time, i.e., $T=M\, \tau$,
with $\tau$ the duration time of a single  $|\alpha\ra$ pulse.
Notice that, $X_{\xi}^{(M)}$ is still a stochastic quantity.
Based on the central-limit-theorem,
the quantum-ensemble average of $X_{\xi}^{(M)}$
and its fluctuation around the average are
\bea
\overline{X}_{\xi}^{(M)} &=& \overline{X}_{\xi}  \,, \nl
\delta X_{\xi}^{(M)} &=& \delta X_{\xi} /\sqrt{M} \,.
\eea
This means that the quantum ensemble averages
of $X_{\xi}^{(M)}$ and $X_{\xi}$ are the same,
but the fluctuation (variance) of $X_{\xi}^{(M)}$
is reduced with the increase of $M$.
In Fig.\ 3, we illustrate this simple behavior.
Especially, in Fig.\ 3(b),
we display the amplified signal $\widetilde{\chi}$
and the squeezing uncertainty of its estimation,
based on $X_{\xi}^{(M)}$,
by increasing the measurement times $M$.
As mentioned above, this is not expensive.
One can just increase the integration time of the output currents,
when using continuous laser beam for the measurement.
This is very reasonable for the case of {\it static parameter} estimation,
but may suffer some limitations on the integration time $T$,
if one considers to detect
time-dependent signal, such as the gravitational waves.
In short, increasing $M$ can just overcome the difficulty
owing to the reduced photon numbers,
which has caused controversial debates
in literature \cite{Nish12,Ked12,Jor14,Li20,Tana13,FC14,Kne14}.  
Moreover, as to be analyzed in the following,
the amplified phase shift itself, has other important technical advantages.

\begin{figure}
  \centering
  \includegraphics[scale=0.6]{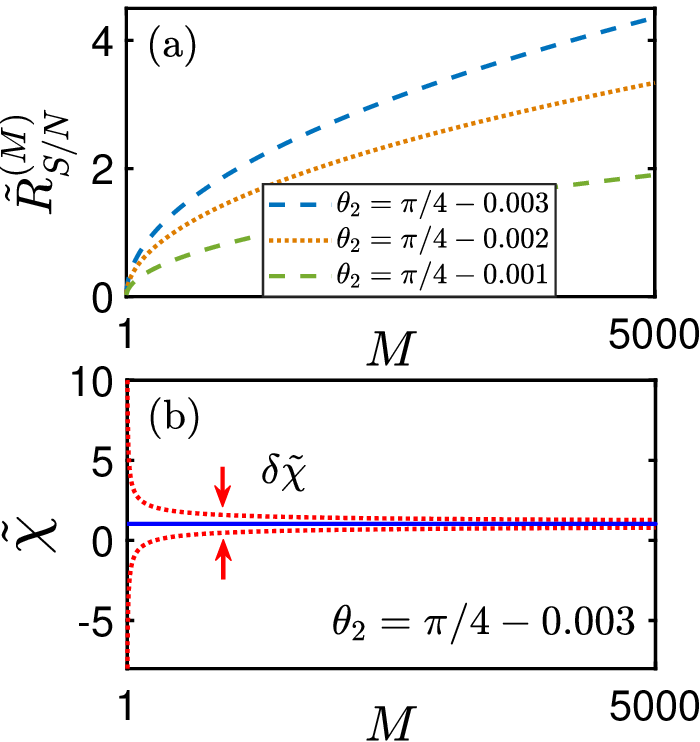}
  \caption{
Using the average result of many times ($M$ times) measurement
to estimate $\widetilde{\chi}$.
In (a), sensitivity (signal-to-noise ratio, SNR)
{\it versus} the measurement times;
and in (b), a more explicit illustration
for the suppression of the estimate uncertainty by increasing $M$,
taking $\theta_2=\pi/4-0.003$ as an example
(which leads to the amplification $\widetilde{\chi}/\chi=10^2$).
Notice that the $M$ times measurement
(using $M$ laser pulses of the coherent state $|\alpha\ra$)
is not expensive, which just corresponds to
increasing the integrating time of the output currents,
when using continuous laser beam for the measurement.
This is in particular reasonable for {\it static parameter} estimation,
which can overcome the difficulty of
reduced photon numbers (thus an increase of shot noise), caused by postselection;
and importantly, which can bring technical advantages in practice,
as successfully demonstrated in the pioneering experiment \cite{Kwi08}.
Parameters assumed in the simulation: $N=100$, and $\chi=10^{-2}$.     }
\end{figure}

\vspace{0.1cm}
{\flushleft\it Technical Advantages}.---
We may first point out that, applying the strategy of PSA,
the amplified phase shift is more tolerant to some technical errors.
For instance, in the homodyne quadrature measurement,
in order to make the result most sensitive to the phase shift under estimation,
we need to set the phase of the LO light as $\xi = \pi/2 + \lambda$,
with $\lambda$ the phase of the incident coherent state $|\alpha\ra$.
Let us consider a modulation error $\delta$,
say, $\xi = \pi/2 + \lambda + \delta$, which is unavoidable in practice.
Then, we have
\bea
\overline{X}_c &=& \sqrt{I_c}\sin(\chi +\delta)   \,, \nl
\overline{X}_{\xi} &=& \sqrt{I_f}\sin(\widetilde{\chi} +\delta)  \,.
\eea
Here $\overline{X}_c$ corresponds to the result of conventional measurement
(direct measurement of the phase shift in $|\alpha e^{i\chi}\ra$, without postselection),
and $\overline{X}_{\xi}$ corresponds to the result of PSA measurement.
Just like in \Eq{XAB}, we use $\widetilde{\chi}$
denoting both $\widetilde{\chi}_A$ and $\widetilde{\chi}_B$, in a unified way.
It is clear that, in the PSA measurement,
the amplified phase shift $\widetilde{\chi}$
is more tolerant to the modulation error $\delta$,
especially, in the case of measuring very small $\chi$.

Next, let us consider another technical issue,
say, the {\it effect of saturation} of photo-detectors \cite {Lun17,ZLJ20}.
The output currents of the two photo-detectors
in the homodyne quadrature measurement can be expressed as
\bea
I_j = k_{\rm max} \left( 1- e^{-N_j/N_{\rm sat}}  \right) \,,
\eea
where $k_{\rm max}$ is the coefficient of opto-electric conversion,
$N_j$ is the number of photons arriving at the $j_{\rm th}$ detector ($j=1$ and 2),
and $N_{\rm sat}$ is the threshold photon number associated with the saturation effect.
In homodyne measurement, we have
$N_1=|\beta + i\alpha_f|^2/2$ and $N_2=|\beta - i\alpha_f|^2/2$.
If $N_1$ and $N_2$ are much smaller than $N_{\rm sat}$,
the output result (the difference of $I_1$ and $I_2$) is
\bea\label{dI}
\Delta I = \left( \frac{k_{\rm max}}{N_{\rm sat}}  \right) (N_1-N_2) \,.
\eea
In particular, after rescaling this $\Delta I$
by the strength of the LO light, $2 |\beta|$, we have
\bea\label{X1}
\overline{X}_{\xi} = \left( \frac{k_{\rm max}}{N_{\rm sat}}  \right)
|\alpha_f| \, \sin\widetilde{\chi}  \,.
\eea
Since the factor $k_{\rm max} / N_{\rm sat}$
can be determined in advance via homodyne measurement
of a weak light--not causing saturation--with known strength and phase,
one can extract the amplified phase shift simply through
\bea\label{ti-chi}
\widetilde{\chi} = \arcsin\left( \overline{X}_{\xi}
\left[ |\alpha_f| (\frac{k_{\rm max}}{N_{\rm sat}}) \right]^{-1} \right) \,,
\eea
once having measured out the field quadrature.
This expression does give the correct result of $\widetilde{\chi}$,
provided that $N_{1,2}\ll N_{\rm sat}$.
However, with the increase of $N_1$ and $N_2$, rather than \Eq{X1},
the homodyne measurement result is given by
\bea\label{X2}
\overline{X}'_{\xi} = \frac{k_{\rm max} }{2|\beta|}
\left( e^{-N_2/N_{\rm sat}} - e^{-N_1/N_{\rm sat}} \right) \,.
\eea
Then, substituting this result into \Eq{ti-chi}
(replacing $\overline{X}_{\xi}$ with $\overline{X}'_{\xi}$),
one will obtain an unreliable $\widetilde{\chi}'$,
and thus fail to infer a correct estimation of $\chi$.

\begin{figure}
  \centering
  \includegraphics[scale=0.5]{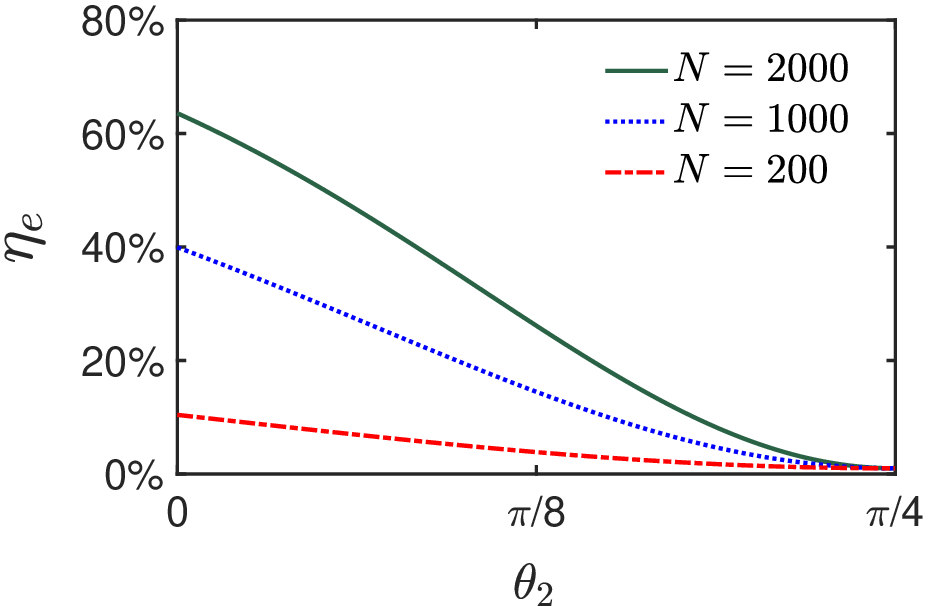}
  \caption{
Estimate error ratio,
$\eta_{\rm e} = |\widetilde{\chi}'-\widetilde{\chi}| / \widetilde{\chi}$,
caused by saturation of the photo-detectors
in the homodyne field-quadrature measurement.
In the regime of weak postselection (small $\theta_2$),
the estimate error becomes more prominent
with the increase of intensity of
the incident light, which is characterized by $N=|\alpha|^2$.
However, even for a strong incident light, e.g., with $N=2000$,
which is much larger than the threshold number $N_{\rm sat}$,
the estimate error can be largely suppressed
along strengthening the postselection.
Parameters assumed in the simulation:
$k_{\rm max}=450$, $N_{\rm sat}=500$, $|\beta|^2=10$, and $\chi=10^{-4}$.      }
\end{figure}

In Fig.\ 4, we numerically illustrate the saturation effect of photo-detectors
on the precision of measurement.
Specifically, we introduce the estimate error ratio as
\bea
\eta_{\rm e} = \frac{|\widetilde{\chi}'-\widetilde{\chi}|}{\widetilde{\chi}}  \,,
\eea
which is vanishing in the regime of $N_1$ and $N_2$ being much smaller than $N_{\rm sat}$.
However, with $N_1$ and $N_2$ approaching to $N_{\rm sat}$
or even being larger than $N_{\rm sat}$,
the saturation effect of photo-detectors will cause serious imprecisions,
as shown in Fig.\ 4.
Indeed, this will happen
if we excessively increase
the intensity of the incident light,
in order to enhance the measurement precision.
In this case, one may adopt the PSA strategy,
which can make the
much weaker light beam (the post-selected photons)
avoid the saturation of the photo-detectors,
while at the same time remaining almost
all the information about the parameter under estimation \cite{Jor14,Li20}.
For instance, as shown in Fig.\ 4, for the strong incident light with $N=2000$
(much larger than the threshold number, $N_{\rm sat}=500$),
along the strengthening of postselection,
the saturation-caused estimation error can be largely suppressed.

\vspace{0.1cm}
{\flushleft\it Summary and Discussion}.---
We have analyzed a postselected amplification (PSA) scheme for phase shift measurement
of optical coherent states when passing through
the Mach-Zehnder-interferometer (MZI).
This work was motivated by the studies in Refs.\ \cite{Ra17,Sab17,Sab18}.
There, the phase shift measurement in the same setup has been studied
in connection with probing relativistic gravity effects,
such as estimating the characteristic parameters of black-holes and wormholes.
Beyond Refs.\ \cite{Ra17,Sab17,Sab18},
owing to intentionally applying the strategy of postselection,
we obtained amplified phase shift,
which can be extracted out from
the field-quadrature measurement in the dark port of the MZI.
We also evaluate the performance quality of the proposed scheme,
and analyze the technical advantages
by considering possible errors in the quadrature measurement.

In present work, we have highlighted that the which-path states of the MZI
cannot be described as two-level sub-system states
entangled with two optical coherent states.
Thus, the postselection cannot generate
quantum superposition of two coherent states,
which has been shown to be capable of enhancing
the precision scaling with the photon numbers \cite{XQLi22,XQLi24}.
This implies that,
for the quadratic coupling studied in Refs.\ \cite{Ra17,Sab17,Sab18},
the postselection cannot enhance the scaling from $1/N^{1.5}$ to $1/N^2$ \cite{XQLi24}.
However, the postselection is expected to be able to amplify the phase change
in the nonlinear MZI, and should hold technical advantages in real measurement.
This will be our forthcoming study along the line of the present work.

In this work, we have also considered in particular
the saturation effect of photo-detectors,
in the context of field quadrature measurement.
This is slightly different from the situation of
transverse deflection measurement of a laser light beam,
studied in Refs.\ \cite{Lun17,ZLJ20}.
There, the pixel noise of camera needs to be considered,
in order to make the postselection scheme
hold technical advantages in the presence of saturation effect.
In our work, we showed that the mere saturation effect
will cause considerable errors in the quadrature measurement,
while the postselection can largely avoid such type of errors,
especially when considering the use of an intense incident light,
in order to enhance the measurement precision.
It is well known that the homodyne quadrature measurement
has been broadly applied in quantum optics.
However, to our knowledge, the effect of saturation
on such type of measurement is not well clarified.
Deeper and detailed investigations for this problem
are required in future research,
including such as how to infer a phase shift
from the field quadrature measurement,
in the presence of saturation effect of detectors.

\vspace{0.5cm}
{\flushleft\it Acknowledgements.}---
This work was supported by the NNSF of China (Nos.\ 11675016, 11974011 \& 61905174).


\end{CJK}
\end{document}